\documentclass[conference]{IEEEtran}
\IEEEoverridecommandlockouts
\usepackage{cite}
\usepackage{amsmath,amssymb,amsfonts}
\usepackage{algorithmic}
\usepackage{graphicx}
\usepackage{textcomp}
\usepackage{xcolor}
\def\BibTeX{{\rm B\kern-.05em{\sc i\kern-.025em b}\kern-.08em
    T\kern-.1667em\lower.7ex\hbox{E}\kern-.125emX}}
\begin{document}

\title{Modified Ring-Oscillator Physical Unclonable Function (RO-PUF) based PRBS Generation as a Device Signature in Distributed Brain Implants\\
}

\author{
\IEEEauthorblockN{Ovishake~Sen and Baibhab~Chatterjee,~\IEEEmembership{Member,~IEEE}}
\IEEEauthorblockA{\textit{Department of ECE, University of Florida (UF)}, Gainesville, USA 32611. email: \{ovishake.sen, chatterjee.b\}@ufl.edu}
}


\maketitle

\begin{abstract}
In this paper, we propose and evaluate a method of generating low-cost device signatures for distributed wireless brain implants, using a Pseudo-Random Binary Sequence (PRBS) Generator that utilizes a modified Ring-Oscillator-based Physical Unclonable Function (RO-PUF). The modified RO-PUF's output is used as a seed for the PRBS generator, which creates a multi-bit output that can be mapped to a time-slot when the implant is allowed to communicate with the external world using duty-cycled time-division multiplexing. A $\textbf{9}$-bit PRBS generator is shown in hardware (with a TSMC $\textbf{65}$ nm test chip implementation) that demonstrates $< \textbf{100}$ nW Power consumption in measurement ($\textbf{72}$\% lower power and $\textbf{78}$\% lower area than a traditional $\textbf{9}$-bit RO-PUF implementation), which supports $\textbf{26}$ implants with the probability of time-slot collision being $< \textbf{50}$\%. This potentially creates a pathway for low-cost device signature generation for highly resource-constrained scenarios such as wireless, distributed neural implants.   
\end{abstract}


\vspace{-2mm}
\section{Introduction}
\subsection{Background and Related Work}
Implantable brain-machine interfaces (BMIc) are becoming increasingly popular due to their immense potential in neurobiological applications. These interfaces primarily consist of two components: (1) neural recording for monitoring brain activity and motor/behavioral patterns, as well as (2) neurostimulators that aid in clinical therapy and physiological studies.  Traditional neural interfaces rely on tethered communication for both data transmission and powering to the site of the implant. This wired approach raises concerns about the potential for cortical scarring, gliosis, infection, and cerebrospinal fluid (CSF) leakage. Consequently, a significant amount of research effort is recently aimed at developing wireless neural interfaces to mitigate these risks.

\begin{figure}[htbp]
\centerline{\includegraphics[width=0.9\columnwidth]{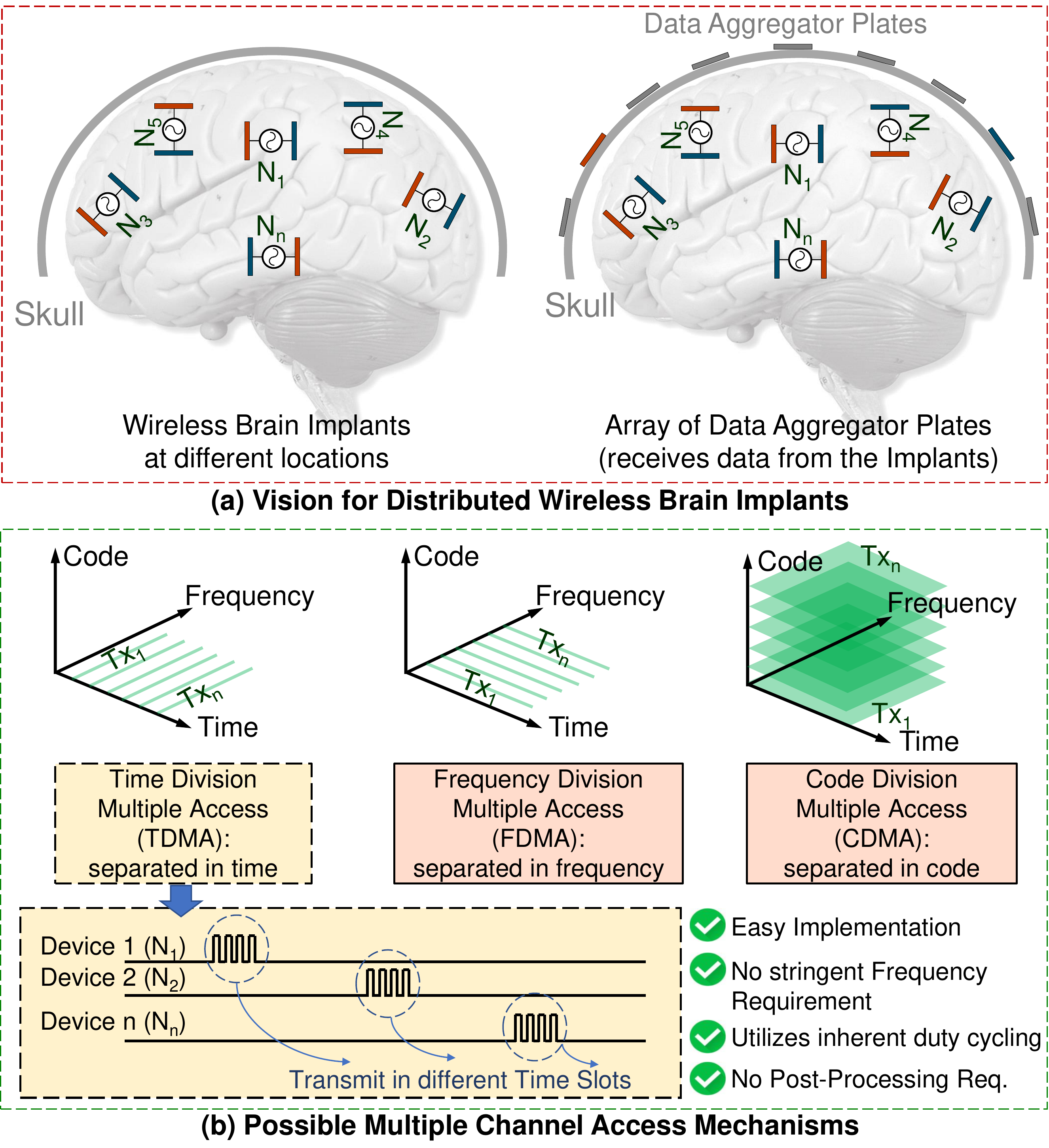}}
\vspace{-2mm}
\caption{(a) Vision for a Distributed Wireless Network of Brain Implants (Node sizes are exaggerated); (b) Possible Multiple Channel Access Schemes for the Implants, and the reason to choose TDMA}
\vspace{-8mm}
\label{fig1}
\end{figure}

\begin{figure*}[htbp]
\centerline{\includegraphics[width=0.9\textwidth]{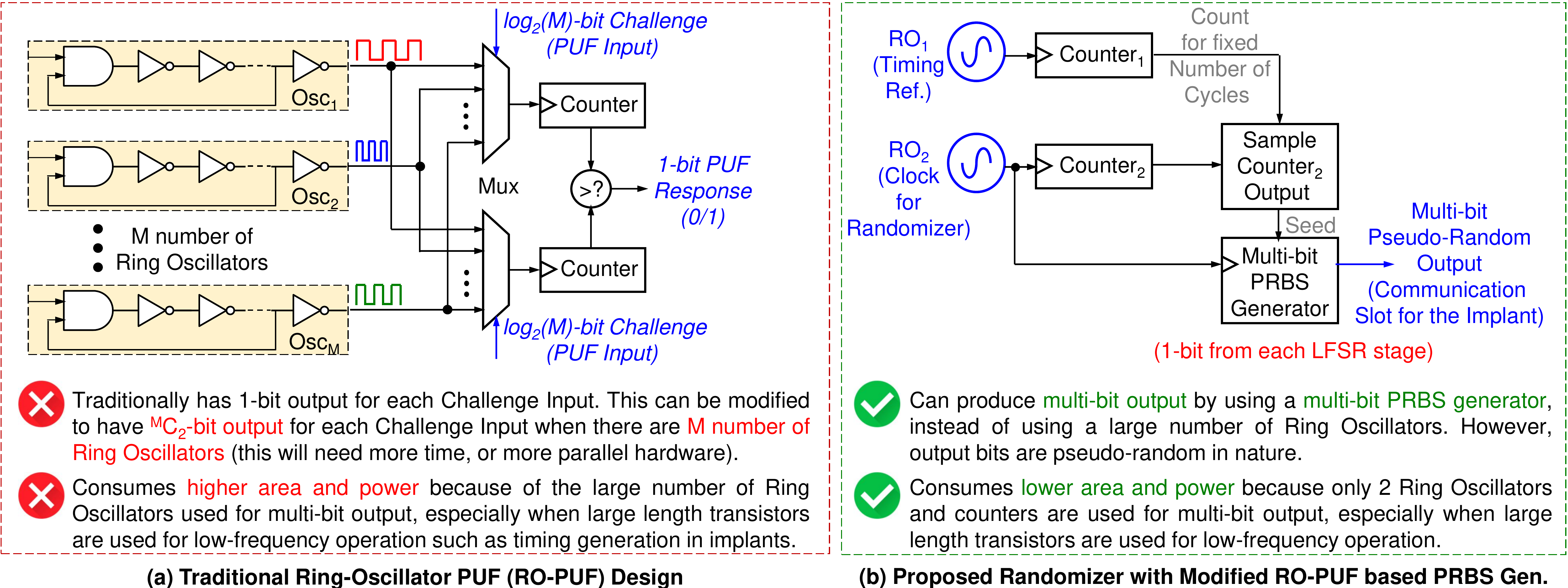}}
\vspace{-2mm}
\caption{(a) Design of a Traditional Ring-Oscillator based PUF (RO-PUF); (b) Proposed Randomizer with Modified RO-PUF based PRBS generation, with the variation in RO frequency being used as a seed to the PRBS Generator.}
\vspace{-5mm}
\label{fig2}
\end{figure*}

However, with any wireless technique, there comes the question of multiple channel access, and recent studies have shown the pros and cons of using techniques such as (1) time-division multiple access (TDMA) \cite{b1, b2, b3}, (2) frequency-division multiple access (FDMA) \cite{b4}, and (3) code-division multiple access (CDMA) \cite{b5}. As shown in Fig. \ref{fig1}, the wireless implants are envisioned to be \emph{sprinkled} throughout the brain, while being connected with the external world (and possibly with each other) for data communication and energy harvesting, thereby creating the concept of an Internet of Body (IoB) \cite{b6, b62}, or in this scenario, an Internet of Brain. The nodes can be separated in time, frequency or in terms of orthogonal encoding. However, in case of FDMA techniques, each node needs to have well-controlled and distinct frequencies, which requires either sophisticated circuitry and/or post-processing. Moreover, for certain systems with interrogators/repeaters, it may require proper frequency synchronization among the implanted devices, the interrogator, and an external hub - thereby increasing the complexity of the overall architecture. Similarly, for synchronous CDMA, clock frequencies of all devices need to be synchronized, while asynchronous CDMA will have scalability issues in the number of nodes supported due to noise aggregation \cite{b1}. Therefore, we utilize a TDMA architecture for multiple access, which inherently takes advantage of the fact that communication with each individual device is most likely duty-cycled due to the consumed energy being larger than the possible harvested energy \cite{b7}. The time slots for TDMA can be either hardcoded for each implant (requires non-volatile memory and post-processing), or can be configured through the downlink communication (complex operation, and hard to scale) from an external device to the implant, or can be found by using physical unclonable function (PUF) embedded in each implant node \cite{b1, b2}. In this work, we use a modified ring-oscillator PUF (RO-PUF) to generate device signatures for TDMA. The RO-PUF is preferred over Arbiter-PUF or SRAM-PUF due to better security and robustness through pairwise comparison \cite{b8}. 

\subsection{Our Contribution}
In this paper, (1) we propose and demonstrate a modified RO-PUF based PRBS generation scheme for device signature generation in resource-constrained implants, which achieve $> 70$\% improvement as compared to traditional RO-PUFs in terms of area and power consumption; (2) we analyze the maximum possible number of implants supported based on the order of the PRBS, and evaluate the effect of the seed of the PRBS to demonstrate the optimum PRBS architecture; (3) finally, we show the proof-of-concept demonstration of a test-chip based on our earlier work \cite{b2}.

\section{Analysis of RO-PUF}
\subsection{Traditional RO-PUFs}
A traditional RO-PUF architecture is shown in Fig. \ref{fig2}(a). For pairwise comparison among $M$-number of ROs, we need to select $2$ out of the $M$ ROs (using $log_2(N)$-bit arbitrary challenge vectors for the $2$ multiplexers (Muxes) shown in Fig. \ref{fig2}(a)), count their frequency output based on $2$ digital counters for the same amount of time, and take a $1$-bit decision based on which counter's value is higher based on manufacturing variations of the two ROs. We can also extend the output to $^MC_2$ bits, with more parallel hardware ($2 \times ^MC_2$ Muxes and counters), or by waiting for many cycles of the operation. Both methods, along with the large number of ROs employed increase the energy and area consumption of the system, which should be avoided for resource-constrained nodes such as brain implants.
\subsection{Proposed $2$-RO PUF with PRBS}
We propose implementing a modified device signature generation scheme with only $2$ ROs (Fig. \ref{fig2}(b)) - RO$_1$ is used as a non-ideal timing reference in the node, while RO$_2$ is used as a clock for the signature generation circuit (randomizer). For a fixed number of cycles of RO$_1$, the frequency of RO$_2$ is counted, which is provided as a seed to a multi-bit pseudo-random binary sequence (PRBS) generator, implemented using linear-feedback shift registers (LFSR), which can generate a pseudo-random multi-bit value based on the seed. The seed is still based on the PUF properties, but the area and power consumption for the randomizer are greatly reduced. 

\begin{figure*}[htbp]
\centerline{\includegraphics[width=\textwidth]{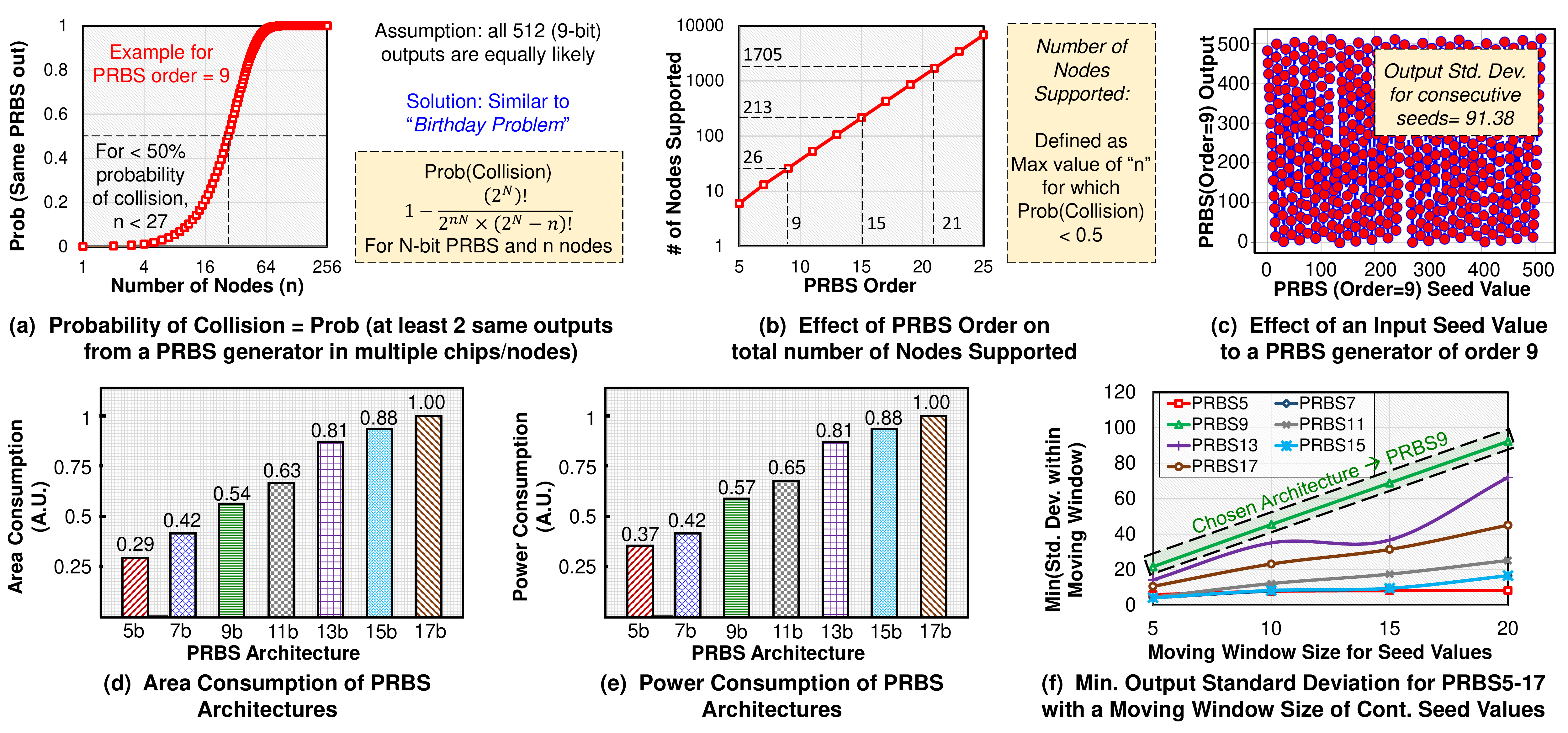}}
\vspace{-2mm}
\caption{Analysis of the Modified RO-PUF based PRBS Generator: (a) Probability of Collision, defined as the probability of at least 2 nodes having the same output from the PRBS Generator; (b) Definition of Total Number of Nodes Supported, and the Effect of PRBS Order on the Number of Nodes Supported; (c) Effect of an Input Seed Value to a PRBS Generator of Order 9; (d) Scaled Area Consumption of PRBS Architectures; (e) Scaled Power Consumption of PRBS Architectures; (f) Minimum Standard Deviation within a moving window of continuous seed values for different PRBS architectures.}
\vspace{-5mm}
\label{fig3}
\end{figure*}

\subsection{Effect of PRBS Order on Possible Number of Implants}
Assuming that the output of the PRBS generator of order $N$ will be able to generate an $N$-bit output which is equally probably between $0$ and $2^N-1$, we can find the possible number of implants that can be supported by an $N$-bit PRBS. Similar to the Birthday Problem \cite{b9}, the probability of the collision (at least two of the $N$-bit values being the same) for $n$ implants can be found as given in Eq. \ref{eq1}.

\begin{equation}
\begin{aligned}
	\text{Prob(Collision)} = 1-\frac{(2^N)!}{2^{nN} \times (2^N-n)!}
\end{aligned}
\label{eq1}
\end{equation}

Fig \ref{fig3}(a) shows the probability of collision vs. the number of nodes for a $9$-bit, PRBS generator of order $9$ (PRBS$9$). For $<50$\% probability of a collision, the number of implants needs to be $<27$. Fig \ref{fig3}(b) presents the number of nodes supported (which we define as the maximum number of nodes for which the probability of collision is $< 0.5$) vs. the PRBS order. As expected, the number of nodes supported increases exponentially as we increase the PRBS order. However, this comes at a cost of area and power.

\subsection{Effect of Seed Value of the PRBS}
Next, we analyze the effect of the initial seed value for the PRBS generation. The seed value is highly dependent on the manufacturing process variations (assuming voltage and temperature remain the same over time which can be achieved using on-chip LDOs and thermal sealing). Interestingly, even with consecutive seeds (very small manufacturing variation), two PRBS outputs can still be far away due to the LFSR polynomial-based bit generation for certain PRBS orders. Fig \ref{fig3}(c) shows the output of PRBS9 with different seeds ($0-511$), which has a standard deviation of $\approx 91.38$ for consecutive seeds. This motivates us to implement the PRBS architecture with the highest min(standard deviation) within a moving window of the seed values. This method promises distinct implant signatures even if the manufacturing process variation is within a small window, resulting in a smaller difference in seed values.

Even if the area and power consumption of PRBS$5$-PRBS$17$ architectures increase almost linearly, as shown in Fig \ref{fig3}(d)-(e), the minimum output standard deviation of these architectures behave very differently within a moving window of the seed values, as shown in Fig \ref{fig3}(f). Because of the large difference in the indices of the generator polynomial (which also means a larger distance between the taps of the LFSR implementation), PRBS$9$ (generator polynomial = $x^9+x^5+1$) has a higher min(standard deviation) than other architectures (for example, PRBS$17$, for which the generator polynomial is $x^{17}+x^{14}+1$ ). The ability to achieve a better standard deviation with lower area and power compels us to  choose PRBS$9$ over some of the higher-order architectures such as PRBS$15$ or PRBS$17$.

\section{Circuit Design and Implementation}
\begin{figure*}[htbp]
\centerline{\includegraphics[width=\textwidth]{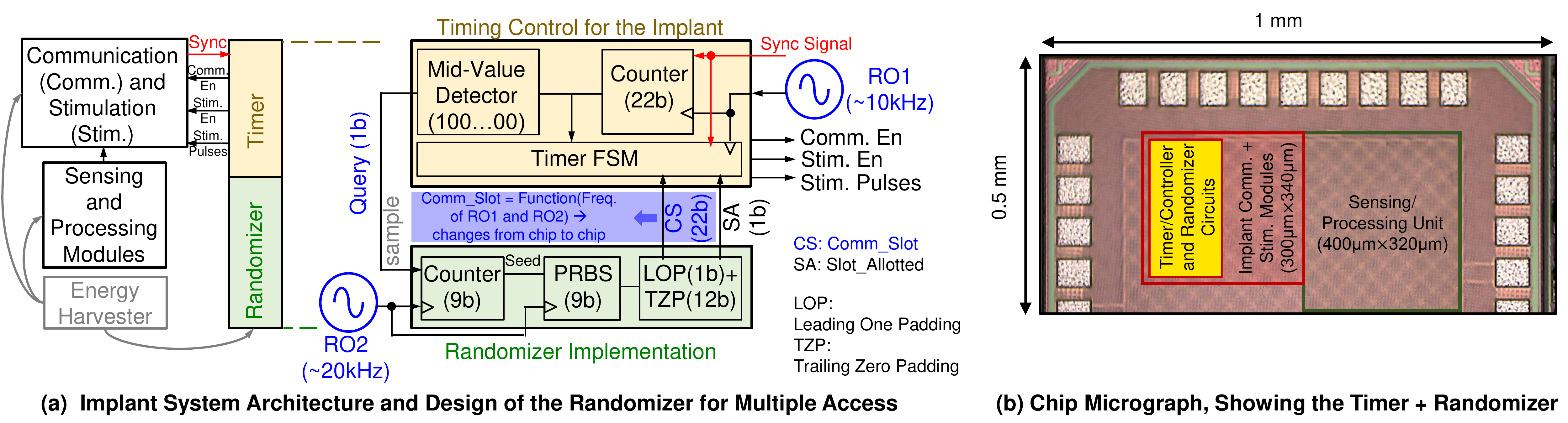}}
\vspace{-2mm}
\caption{(a) Implant System Architecture and the Design of the Modified RO-PUF based Communication/Stimulation Slot Randomizer Unit; (b) Chip Micrograph, showing the Timer and Randomizer Units. Part of the figure is taken from \cite{b2}.}
\vspace{-5mm}
\label{fig4}
\end{figure*}

Based on the above analysis, we have designed and implemented a PRBS$9$ architecture with $2$ ROs as the randomizer, shown in Fig. \ref{fig2}(b). The design details of the entire implant can be found in \cite{b2,b3}, which aimed to explain the details of a new modality of communication in the brain implants called Bi-Phasic Quasistatic Brain Communication (BP-QBC), and did not focus on the full analysis of the device signature generation. Fig. \ref{fig4}(a) shows the simplified block diagram of the implant IC that has been implemented in a $65$ nm CMOS process \cite{b2} and contains an energy harvester, an on-chip sensing and processing module, a communication and stimulation module, along with a timer and randomizer circuit for device signature generation, which is also utilized for communication/stimulation slot selection with TDMA for the implant.

\subsection{Ring Oscillator (RO) Implementation}
RO$_1$ and RO$_2$ for the implant are implemented as $5$-stage, single-ended $10$ kHz and $20$ kHZ oscillators for low area and power. Since these clocks are used for low-frequency timing (and can be non-exact), we don't need to generate MHz of GHz frequencies. However, to generate such low frequencies with only a 5-stage oscillator, we needed to increase the transistor lengths which also helped to reduce the current consumption as well as leakage.
\subsection{PRBS9 based Device Signature and Communication/Stimulation Slot Generator}
The output of RO$_1$ controls the timer Finite State Machine (FSM) for the IC, as well as a $22$-bit counter. A Synchronization (Sync) signal resets the counter and the FSM, which can be broadcasted to all the implants using the downlink (external device to implant) communication channel. When the $22$-bit counter reaches the middle of its full range, a query is made to a separate $9$-bit counter which is part of the randomizer, and runs on the RO$_2$ clock. The $9$-bit counter's output during the query (which is dependent on the manufacturing process dependent frequency difference of RO$_1$ and RO$_2$) is used as a seed to the PRBS generator ($9$-bit). The $9$-bit output of the PRBS generator (device signature) is padded with $1$-bit leading one and  $12b$ trailing zeros to create a $22$-bit Communication Slot (CS) output which is provided to the timer FSM. This works as the start of the communication for the particular implant. The leading one ensures that the selected slot is after raising the query in time domain and before the $22$-bit counter resets. The trailing zeros help the timer to compare the CS value with the $22$-bit counter's output and start the communication protocol. According to the duty-cycled architecture shown in \cite{b2}, the communication is active for $\approx 100$ ms, followed by a $\approx 100$ ms stimulation, with $\approx 100$ ms inactivity in between. This entire cycle repeats after a reconfigurable amount of time ($\approx 100$ s by default), and is controlled by the timer FSM. Fig. \ref{fig4}(b) shows the chip micrograph, along with the area for the timer/randomizer circuit, which is about $0.09$mm$^2$.

\section{Measurement Results}

\begin{figure}[htbp]
\centerline{\includegraphics[width=0.7\columnwidth]{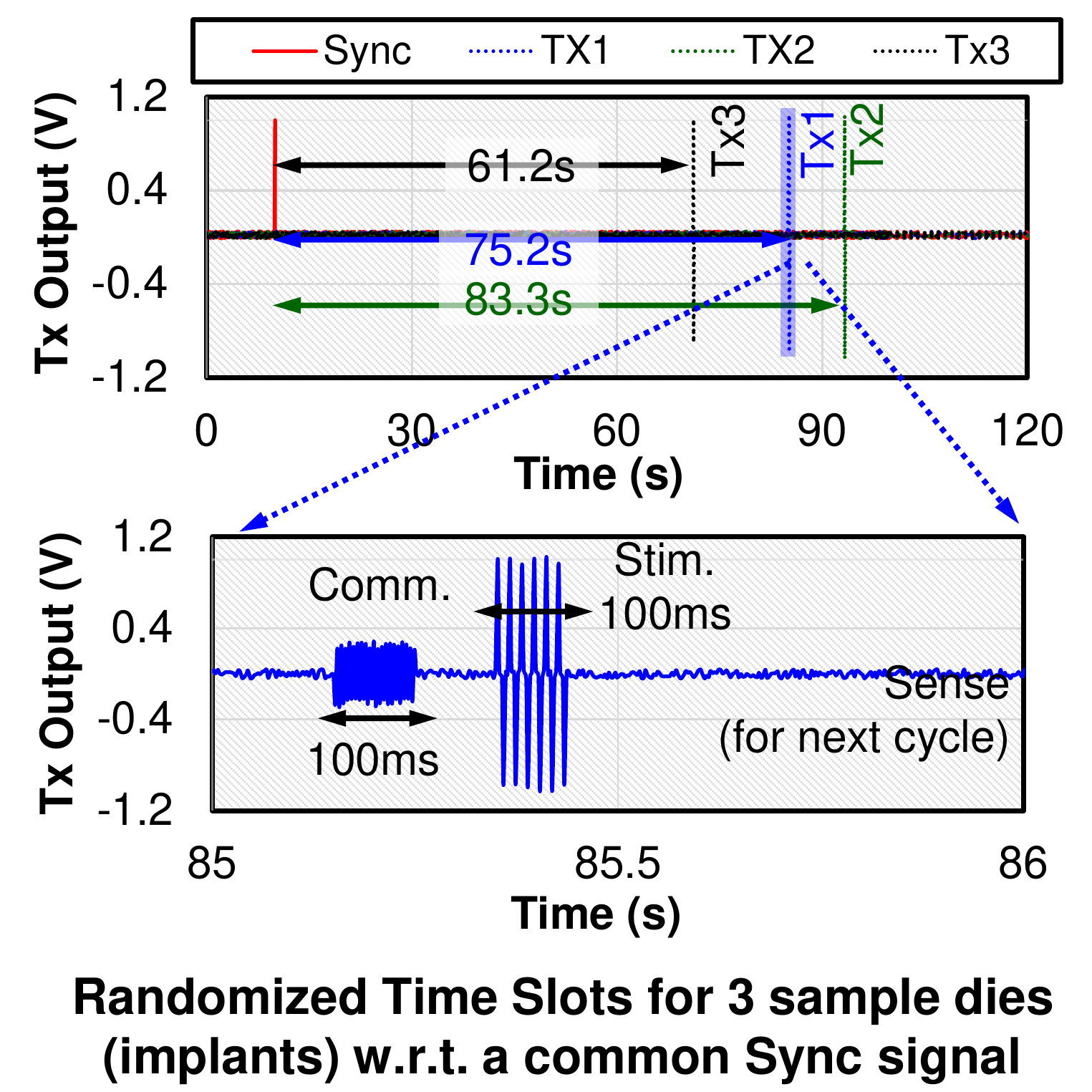}}
\vspace{-2mm}
\caption{Measured Randomizer Time Slots for 3 Sample Dies (Implants) w.r.t. a Common Synchronization Signal. The data is taken from \cite{b2}.}
\vspace{-2mm}
\label{fig5}
\end{figure}

The measurement result for the $65$ nm test-chip is shown in Fig. \ref{fig5}, and is taken from \cite{b2}. Randomized time slots for 3 sample dies are shown after providing a common Sync signal externally in a table-top measurement setup (without implanting the chip). The 3 samples start their communication/stimulation cycles at about $61.2$ s, $75.2$ s and $83.3$ s after the sync signal. The overall power consumption of the timer+randomizer unit is only $81$ nW (as compared to $\approx 300$ nW for a traditional $9$-bit RO PUF) at $0.4$ V Supply. The low-power operation is benefitted from using HVT transistors in the design of the timer and randomizer. The overall area reduction as compared to the traditional $9$-bit RO PUF is about $78$\% for the timer+randomizer, and about $91$\% for only the randomizer. 

\section{Conclusions and Future Work}
 We have analyzed the applicability of a modified RO-PUF based PRBS generator for device signature (as well as communication/stimulation slot) generation in a group of distributed wireless brain implants that utilize TDMA for multiple access. TDMA offers simple architectures, and benefits from the inherent duty-cycled nature of these implants.  The model for the number of maximum supported implants based on the order of PRBS is explained, and the effects of the seed value (which is dependent on the manufacturing process variations) of the PRBS on the standard deviation of the PRBS output is analyzed, to indicate the optimum PRBS architecture. In future work, optimum architectures for other PUF structures (for example, SRAM-PUF and layout-dependent RC-delays) will be explored for resource-constrained implants. 


\section*{Acknowledgment}

The authors thank Dr. Shreyas Sen, Purdue University for feedback. The authors also thank TSMC and Muse Semiconductor for MPW runs and Shuttle Services.

\vspace{12pt}

\end{document}